\documentclass{iopconfser}

\usepackage{amsmath}
\usepackage{amssymb}
\usepackage{graphicx}
\usepackage{hyperref}
\usepackage{xspace}
\usepackage{orcidlink}
\usepackage{lmodern}
\usepackage{anyfontsize}

\newcommand{\ICRR}{Institute for Cosmic Ray Research, The University of Tokyo, 5-1-5 Kashiwanoha, Kashiwa, Chiba 277-8582, Japan}
\newcommand{\IUCAA}{Inter-University Centre for Astronomy and Astrophysics, Post Bag 4, Ganeshkhind, Pune 411 007, India}
\newcommand{\IPMU}{Kavli Institute for the Physics and Mathematics of the Universe (WPI), The University of Tokyo, 5-1-5, Kashiwanoha, 277-8583, Japan}

\begin{document}

\title{Constraining the Hubble Constant using Cross-Correlation of Gravitational Wave Events with Flux-Limited Galaxy Catalog}

\author{Tathagata Ghosh~\orcidlink{0000-0001-9848-9905}$^{1}$ and Surhud More~\orcidlink{0000-0002-2986-2371}$^{2, 3}$}

\affil{$^1$\ICRR} 
\vspace{12pt}

\affil{$^2$\IUCAA} 
\vspace{12pt}

\affil{$^3$\IPMU}
\vspace{12pt}

\begin{abstract}

Gravitational waves (GWs) from the compact binary coalescence provide direct measurement of the luminosity distance to the event. However, unlike binary neutron stars, redshift information is not available from GW observations of binary black holes. Consequently, independent redshift measurements of such GW events are necessary to measure $H_0$. In this study, we demonstrate a novel Bayesian formalism to infer $H_0$ utilizing the $3$D cross-correlation of GW events with galaxies from flux-limited catalog in configuration space. We demonstrate the efficacy of our method with $300$ simulated GW events distributed within $1$ Gpc in colored Gaussian noise of Advanced LIGO and Advanced Virgo detectors operating at O4 sensitivity. We show that such measurements can constrain the Hubble constant with a precision of $\sim 9 \%$ ($90\%$ highest density interval). We highlight the potential improvements that need to be accounted for in further studies before the method can be applied to real data.

\end{abstract}

\section{Introduction}

The discovery of gravitational wave (GW) events has provided an independent probe to study the expansion history of the Universe. Except for the first and only multimessenger event, GW170817~\cite{LIGOScientific:2017vwq}, current measurements primarily rely on binary black hole (BBH) mergers~\cite{KAGRA:2021vkt}. These events, lacking electromagnetic (EM) counterparts, are referred to as \textit{dark sirens}. As the unique identification of the host galaxy for such BBHs is not currently possible due to large sky localization errors, there are several approaches to infer the Hubble constant from such dark sirens. 
The current state-of-the-art methods~\cite{Gray:2021sew, Mastrogiovanni:2021wsd} adopted to estimate $H_{0}$ include the statistical host identification method while allowing the fitting of an astrophysics-motivated source frame mass model to the observed mass distribution of GW events.
However, this method does not utilize the information on the spatial clustering of galaxies explicitly. Both the GW source population and the galaxy population are tracers of the same underlying large-scale structure in the Universe. So, GW sources share the same large-scale structure as galaxies at their redshifts and, hence, are correlated to each other. Therefore, one can expect that the cross-correlation function between galaxies with known redshift and GW sources would be nonzero at the true redshift of the GW sources and, thus, in conjunction with the luminosity distances of the GW events, can be used to constrain the cosmological parameters such as the Hubble constant~\cite{Oguri:2016dgk, Bera:2020jhx, Ghosh:2023ksl}. In this work, we extend the cross-correlation method proposed in Ref.~\cite{Ghosh:2023ksl} to infer $H_{0}$ from \textit{individual} GW events in order to utilize the cross-correlation between these GW events and a flux-limited galaxy catalog. 

\section{Method and Simulation}

In this study, we follow the novel Bayesian framework from Ref.~\cite{Ghosh:2023ksl}. This approach models the galaxy overdensity for each GW event while also accounting for its volume uncertainty region when estimating the Hubble constant through this cross-correlation.
Ref.~\cite{Ghosh:2023ksl} demonstrates the efficacy of the cross-correlation method for inferring the Hubble constant from individual GW events as a proof-of-concept. The mock galaxy catalog was assumed to be volume-limited with no limit on the galaxy fluxes. In contrast, real galaxy catalogs are flux-limited. 
This study examines how using a flux-limited galaxy catalog affects $H_{0}$ estimation. We also compare $H_{0}$ measurements derived from both complete and flux-limited galaxy catalogs.
We first outline the procedure to construct such a galaxy catalog. After that, we compare $H_{0}$ estimates obtained from both types of catalogs in Sec.~\ref{sec:result}.
In this study, we only consider $\theta_{\rm max}=0.02$ radian to compute the $H_{0}$ posterior for both the galaxy catalogs. This choice is based on our earlier analysis in Ref.~\cite{Ghosh:2023ksl}, which indicates that $\theta_{\rm max} = 0.02$ radian is the optimal one. However, further detailed studies may be needed to confirm this choice.

We consider the same galaxy catalog as described in Ref.~\cite{Ghosh:2023ksl}. We calculate the apparent magnitudes of the galaxies to construct a flux-limited catalog out of this complete galaxy catalog.
To achieve that, we calculate the absolute magnitude $M$ of the galaxies using the $M-v_{\rm max}$ relation from Ref.~\cite{2011ApJ...742...16T}. To make the galaxy catalog realistic, we further add random Gaussian errors with mean $0$ and standard deviation $0.5$~\cite{2011ApJ...742...16T} to these estimated absolute magnitudes of the galaxies to account for the scatter in the $M-v_{\rm max}$ relation. The absolute magnitudes thus obtained are now used along with their luminosity distances (calculated based on the cosmology used in the BigMDPL simulation, as detailed in Sec. III of Ref.~\cite{Ghosh:2023ksl}) to compute the corresponding apparent magnitudes.
Though the GW events are randomly assigned to the complete galaxy catalog, the galaxies with apparent magnitudes of $m \leq 17.77$, similar to spectroscopic galaxies from the Sloan Digital Sky Survey, are utilized for inferring the Hubble constant. We also estimate the Hubble constant using the complete galaxy catalog for a comparison of the statistical power for constraining $H_{0}$ between the two galaxy catalogs. 

\begin{figure}
    \centering
    \includegraphics[scale=0.52]{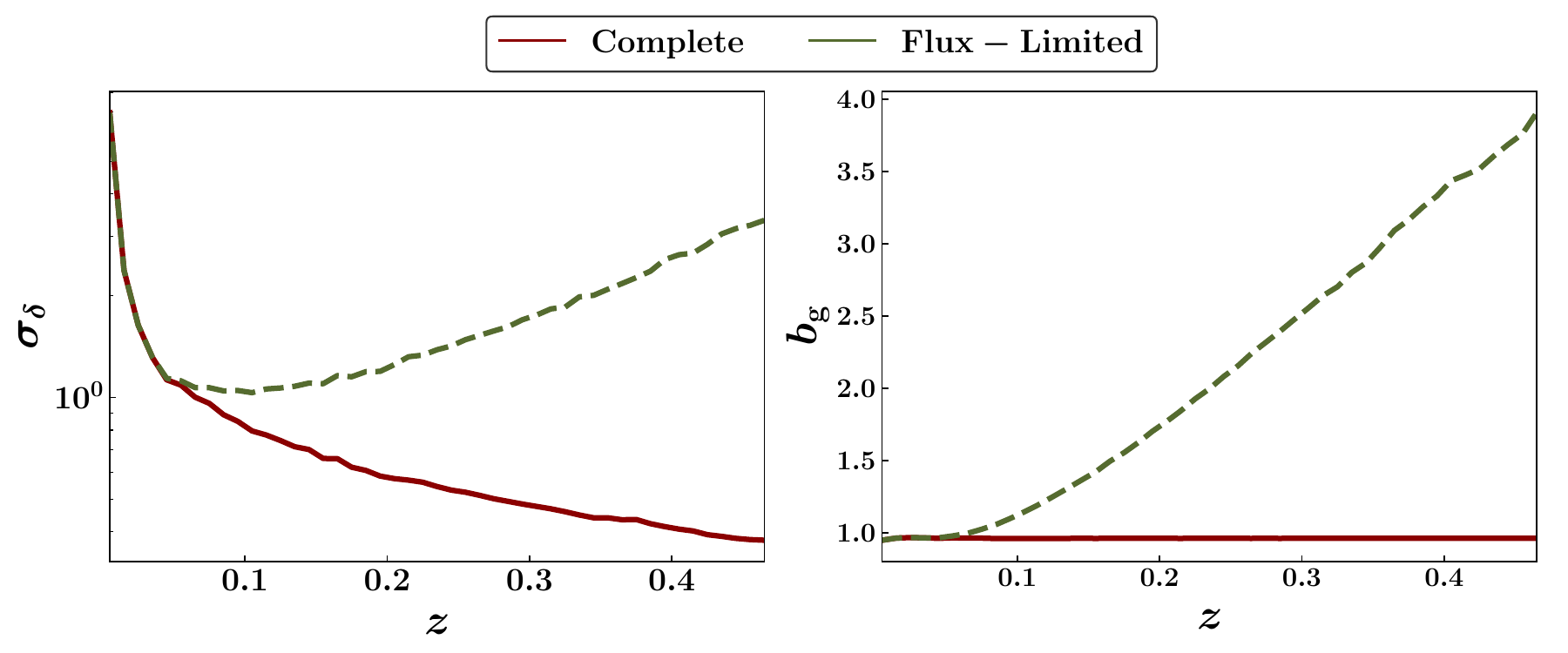}
    \caption{\textit{Left Panel:} Comparison of overdensity fluctuation $\sigma_{\delta}$ for the complete and flux-limited galaxy catalog considering $\theta_{\rm max}=0.02$ radian. 
    \textit{Right Panel:} Galaxy bias, $b_{g}$ as a function of redshift for the complete and flux-limited galaxy catalogs. The solid lines in both panels correspond to the complete galaxy catalog from Ref.~\cite{Ghosh:2023ksl}, and the dashed lines correspond to the flux-limited catalog constructed from it in this work.}
    \label{fig:sigma_bias_compare}
\end{figure}

The corresponding overdensity fluctuations for both galaxy catalogs are shown in the left panel of Fig.~\ref{fig:sigma_bias_compare}. We follow the same process described in Sec. III of Ref.~\cite{Ghosh:2023ksl} to calculate overdensity fluctuations from $10^{6}$ random lines-of-sight within the galaxy catalogs. For the complete galaxy catalog, the overdensity fluctuation is expected to decrease with redshift because the cosmic volume increases with redshift, accommodating more galaxies. At low redshifts, all galaxies are visible in the flux-limited galaxy catalog. However, as redshift increases, only bright galaxies can be seen at higher redshift due to the decrease in their fluxes, which in turn increases the variance of the overdensity fluctuations. Therefore, while the variance of the overdensity fluctuations for the flux-limited galaxy catalog initially follows those of the complete galaxy catalog at lower redshifts, they start to increase after a particular redshift. 

\section{Results} \label{sec:result}

We apply the method mentioned as implemented in Ref.~\cite{Ghosh:2023ksl} to infer the Hubble constant using both the complete and flux-limited galaxy catalog. It is important to note that the galaxy bias factor needs to be accounted for appropriately while considering the flux-limited galaxy catalog. For the complete galaxy catalog, the galaxy bias factor does not evolve with redshift due to its completeness through the redshift range. However, the galaxy bias factor for the flux-limited evolves with redshift. In particular, this galaxy bias dependence on redshift can be inferred from the auto-correlation of the galaxies in the catalog as a function of redshift.

In our case, we calculate the galaxy bias directly from the dark matter halo catalog. As detailed in Ref.~\cite{Ghosh:2023ksl}, the galaxies are assumed to be assigned to all the central dark matter halos, allowing us to determine the mass of the dark matter halo associated with each galaxy. The galaxy bias at the redshift of galaxies is calculated following Eq.~(2) of Ref.~\cite{Ghosh:2023ksl}, where the bias for a given mass of dark matter halo is taken from Ref.~\cite{Tinker:2006nn}. Note that we use the relation between bias and halo mass at the redshift corresponding to the simulation.  The redshift dependence of the galaxy bias for the flux-limited catalog is shown in the right panel of Fig.~\ref{fig:sigma_bias_compare} and is purely a result of the changing absolute luminosity of galaxies in a flux-limited catalog as a function of redshift. As the number of galaxies being observed decreases with redshift for flux-limited galaxy catalog, the galaxy bias is expected to increase. In real data, we lack information on the true halo mass distribution. However, measurements of galaxy clustering amongst themselves would offer an empirical expectation for the bias and its evolution with redshift.

\begin{figure}
    \centering
    \includegraphics[scale=0.57]{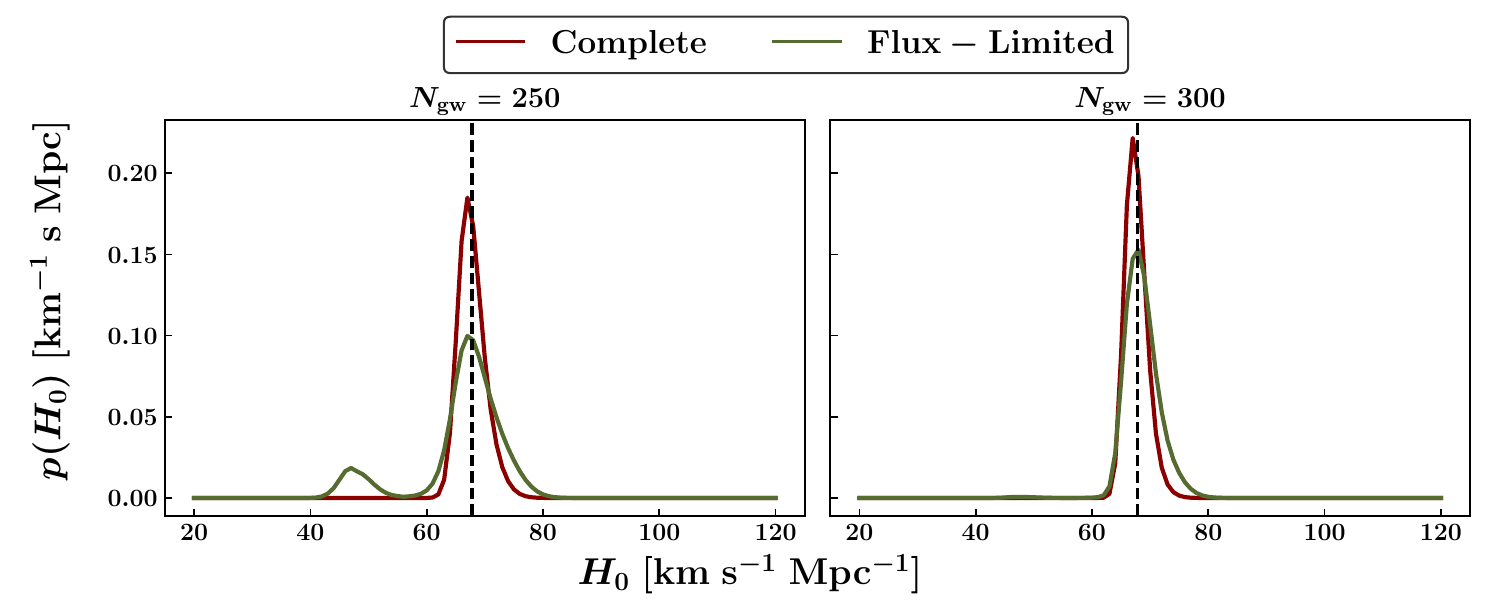}
    \caption{Comparison of $H_{0}$ posteriors inferred using the complete and flux-limited galaxy catalogs for different numbers of GW events as indicated at the top of the respective plots~\cite{Ghosh:2023ksl}.}
    \label{fig:pH0_mcut_compare}
\end{figure}

The posteriors of the Hubble constant inferred by considering both the complete and flux-limited galaxy catalogs are shown in Fig.~\ref{fig:pH0_mcut_compare}. Though $H_{0}$ posterior calculated by cross-correlating $250$ GW events with the complete galaxy catalog is quite precise, $H_{0}$ posterior calculated by cross-correlating $250$ GW events with the flux-limited galaxy catalog is relatively significantly broad. This feature is expected as the complete galaxy catalog provides much large-scale information at high redshift due to the presence of more galaxies. However, with the increase in the GW events, the statistical precision of the estimation of $H_{0}$ for the flux-limited galaxy catalog increases.

\section{Summary}

In this study, we successfully demonstrated the inference of the Hubble constant from individual GW events by cross-correlating with the flux-limited galaxy catalog as a proof-of-concept. This method is useful even if the mergers do not have any EM observations and can supplement the inference of the Hubble constant from BNS mergers with detected EM counterparts. Although we have considered a flux-limited galaxy catalog here, we have not accounted for varying depths in different directions, which can be a straightforward extension of our current approach.  
We have also considered the idealistic scenario where the GW events are the unbiased tracers of the underlying matter distribution. In reality, the GW events and the galaxies trace the large-scale structure with a redshift-dependent bias. This redshift-dependent bias could arise from the evolving way in which the gravitational wave events populate dark matter halos or due to selection effects. These effects need to be parameterized and marginalized to constrain the cosmological parameters from real data.
We defer the application of our method to galaxy catalogs, which include these imperfections, to future work.

\section*{Acknowledgments}

T.G. acknowledges the use of the IUCAA LDG cluster Sarathi for computations and support from JSPS Grant-in-Aid for Transformative Research Areas (A) No. 23H04893.

\bibliography{reference}

\end{document}